\documentstyle[aps,prl,epsfig,twocolumn]{revtex}
\begin{document}
\draft
\twocolumn[\hsize\textwidth\columnwidth\hsize\csname @twocolumnfalse\endcsname

\title{Impact of magnetic frustration on the Mott transition\\ within a
slave-boson mean-field theory}
\author{Imseok Yang, Ekkehard Lange, and Gabriel Kotliar}
\address{Serin Physics Laboratory, Rutgers University, 136
Frelinghuysen Road, Piscataway, New Jersey 08854, USA
}
\maketitle
\begin{abstract}

We investigate the paramagnetic-metal-to-antiferromagnetic-metal and  
antiferromagnetic-metal-to-antiferromagnetic-insulator transitions
using a slave-boson mean-field theory.  To this effect, we discuss the
ground state of the half-filled Hubbard model as a function of
$t^{\prime}/t$ and correlation strength $U$, where $t$ and
$t^{\prime}$ are the hopping amplitudes between nearest and
next-nearest neighbors, respectively.  The metal-insulator transition
at a critical $U_{\mbox{\scriptsize MIT}}$ is of second order for
small levels of magnetic frustration, $t'/t<0.06$, and of first order
for large ones, $t'/t>0.06$. The insulator is always
antiferromagnetically ordered, while the metal exhibits a second-order 
transition from a paramagnetic to an antiferromagnetic state up to
$t'/t=0.14$, as $U$ is increased. We also contrast these findings with
what we obtain in Hartree-Fock approximation.

\end{abstract}

]

The correlation-driven metal-insulator transition, or Mott transition
\cite{Mott:1961}, observed in materials such as V$_2$O$_3$ 
\cite{Carter:1993,Imada:1998} and NiS$_{2-x}$Se$_x$
\cite{Imada:1998,Gautier:1975,Sudo:1992}, is a nonperturbative
problem usually tackled within the Hubbard model of strongly
correlated electrons. This model describes itinerant electrons subject
to an on-site repulsion $U$ comparable or greater than the bare
bandwidth $2D$. 

In an early work, Brinkman and Rice \cite{Brinkman:1970} investigated
the Mott transition from the metallic side using Gutzwiller's
variational scheme \cite{Gutzwiller:1963,Gutzwiller:1965}. In this
approximation, the metal is described as a strongly renormalized Fermi
liquid. A low-energy scale $ZD$ ($Z$ is the quasiparticle residue)
collapses linearly in $U$ as the Mott transition, occuring at
a critical $U_{\mbox{\scriptsize BR}}$, is approached from the
metallic side. $ZD$ is a measure for the renormalized Fermi energy. 

We investigate the implications of antiferromagnetic long-range order
on either side of the Mott transition. To this effect, we introduce
magnetic frustration, which helps stabilize an antiferromagnetic
metallic phase for not too large levels of frustration and causes the
insulating side to favor antiferromagnetic long-range order
\cite{Rozenberg:1994,Lin:1987,Duffy:1997,Chitra:1998}. We determine
the phase diagram and the orders of the transitions in mean-field
theory using a slave-boson technique. Our work complements previous
studies \cite{Lin:1987,Duffy:1997,Chitra:1998} and adds some
analytical insights on the interplay of electron-electron correlations
and magnetism.

The single-band Hubbard model is given by
\begin{equation}
 \hat{H}=-\sum_{ij\sigma}t_{ij}c_{i\sigma}^{+}c_{j\sigma}
 +U\sum_i \hat{n}_{i\uparrow}\hat{n}_{i\downarrow},
\label{Hubbard}
\end{equation}
where we take the amplitudes $t_{ij}$ to be nonzero only between
nearest and next-nearest neighbors, in which cases they equal $t$ and
$t^{\prime}$, respectively. $c_{i\sigma}^{+}$ and $c_{i\sigma}$ are
creation and annihilation operators for an electron of spin $\sigma$
at site $i$, and $\hat{n}_{i\sigma}\equiv c_{i\sigma}^{+}c_{i\sigma}$.
In this work, we consider the two-dimensional cubic lattice and
restrict ourselves to half filling and zero temperature. 

We use a slave-boson method \cite{Kotliar:1986} in which the
approximation of Gutzwiller, Brinkman, and Rice is recovered on the
saddle-point level, but which at the same time is open to various
generalizations such as the inclusion of antiferromagnetic long-range
order. The slave-boson method yields a ground-state energy in good
agreement with quantum Monte Carlo simulations including
antiferromagnetic order \cite{Lilly:1990} and spiral states
\cite{Fresard:1991}, or with exact diagonalization studies
\cite{Fresard:1992}. It has also been used to determine a magnetic
phase diagram \cite{Moeller:1993} and to go beyond the Hartree-Fock
approximation in problems involving complicated spatial structures
such as stripes \cite{Seibold:1998}.

In the Kotliar-Ruckenstein approach, two aspects of a physical
electron are separated: that it is a {\em fermion} and that it affects
the {\em occupancy} of some site. The first aspect is taken into
account by a fermionic field $f_{i\sigma}$, while the possible
occupancies of the sites are described by bosonic fields: $e_i$
describes empty, $p_{i\sigma}$ singly occupied, and $d_i$ doubly
occupied sites. The physical electron field is represented as
$c_{i\sigma}=\tilde{z}_{i\sigma}f_{i\sigma}$ with 
$\tilde{z}_{i\sigma}=
	(1-p^+_{i\sigma}p_{i\sigma}-d_i^+d_i)^{-1/2}
	(e_i^+p_{i\sigma}+p^+_{i-\sigma}d^+_i)
	(1-e_i^+e_i-p^+_{i-\sigma}p_{i-\sigma})^{-1/2}$,
while appropriate constraints eliminate unphysical states
\cite{Kotliar:1986}. Thus, the problem posed by the Hubbard
interaction is shifted to that of keeping track of the backflow of
bosonic excitations,
$\tilde{z}_{i\sigma}^+\tilde{z}_{j\sigma}$, accompanying
the itinerant fermions, $f_{i\sigma}^+f_{j\sigma}$. 

Proceeding along the lines of Ref.\ \cite{Kotliar:1986}, we first
set up the functional-integral representation of the Hubbard model in
terms of the above-mentioned auxiliary fields, integrate out the
fermions, and solve the remaining problem in the saddle-point
approximation. To describe antiferromagnetism, we divide the lattice
into two sublattices, $A$ and $B$, and look for solutions satisfying
the following relations between the sublattice Bose fields: $e_B=e_A$,
$p_{B\sigma}=p_{A-\sigma}$, $d_B=d_A$, and
$m=p_{A\uparrow}^2-p_{A\downarrow}^2$, where $m$ is the staggered
magnetization. For our result, we need the dispersion relations of the
renormalized quasiparticle bands,
\begin{equation}
 \epsilon_{\vec{k}\eta}[X]=-4t^{\prime}\cos k_x\cos k_y
	+\eta t\sqrt{X^2+4(\cos k_x+\cos k_y)^2} 
\label{dispersions},
\end{equation}
where the lattice spacing has been set equal to one, $X$ is some
dynamically generated staggered magnetic field, and $\eta=\pm1$. The
equations for the density per site (which at half filling is equal to
one) and the staggered magnetization,
\begin{eqnarray}
 1&=&\frac{1}{N}\sum_{\vec{k}\eta}
	f(\epsilon_{\vec{k}\eta}[X]-\tilde{\mu}) 
\label{density},\\
 m(X,\tilde{\mu})&=&\frac{1}{N}\sum_{\vec{k}}
     	\frac{X}{\sqrt{X^2+4(\cos k_x+\cos k_y)^2}}
\nonumber\\
	&&\times\left[f(\epsilon_{\vec{k}-}[X]-\tilde{\mu})
	-f(\epsilon_{\vec{k}+}[X]-\tilde{\mu})\right],
\label{magnetization}
\end{eqnarray}
can be solved unambiguously for $X$ and the effective chemical
potential $\tilde{\mu}$, to yield functions $\tilde{\mu}(m)$ and
$X(m)$. In Eqs.\ (\ref{density}) and (\ref{magnetization}), $N$ is the
total number of lattice sites, the sum is over the first Brillouin
zone, and $f(\epsilon)=\Theta(-\epsilon)$ is the Fermi function at
zero temperature. From the mean-field equation 
\begin{equation}
 \frac{\partial q(m,d^2)}{\partial d^2}K(m)+U=0,
\label{dsquareofm}
\end{equation}
where the functions $q(m,d^2)$ and $K(m)$ are given by
\begin{eqnarray}
 q(m,d^2)&=&\frac{4d^2}{1-m^2}
	\left[1-2d^2+\sqrt{(1-2d^2)^2-m^2}\right]
\label{q-renormalization},\\
 K(m)&\equiv&\frac{1}{N}\sum_{\vec{k}\eta}
  \epsilon_{\vec{k}\eta}[X(m)]
  f(\epsilon_{\vec{k}\eta}[X(m)]-\tilde{\mu}(m))
\nonumber\\
     &&+tmX(m),
\label{kinetic}
\end{eqnarray}
we obtain the average portion of doubly occupied sites as a function
of the staggered magnetization, $d^2(m)$. This function along with
Eqs.\ (\ref{q-renormalization}) and (\ref{kinetic}) allow to write
the ground-state energy per site as a function of the staggered
magnetization as
\begin{equation}
 e(m)=q(m)K(m)+Ud^2(m).
\label{Landau-function}
\end{equation}

This result has an intuitive interpretation: $K(m)$ is the kinetic
energy of noninteracting lattice fermions with nearest and
next-nearest neighbor hopping, subject to an internal staggered
magnetic field $tX(m)$. The renormalization factor $q(m)$ accounts for
the reduction of the hopping amplitudes due to the local correlations
and is characteristic of the Gutzwiller approximation. In our scheme,
$q$ arises from the expectation value 
$\langle\tilde{z}_{i\sigma}^+\tilde{z}_{j\sigma}\rangle$ and thus
represents the average effect of the backflowing slave bosons. The
second term in Eq.\ (\ref{Landau-function}) is the contribution of 
the Hubbard interaction to the energy. 

For $U=0$, Eqs.\ (\ref{dsquareofm}) and (\ref{q-renormalization})
yield $d^2=(1-m^2)/4$ and $q=1$, so $\epsilon(m)=K(m)$.

For $t^{\prime}=m=0$, Eqs.\ (\ref{dispersions})-(\ref{kinetic}) imply:
$X=\tilde{\mu}=0$;
$K=2\int_{-\infty}^0d\epsilon\,D_0(\epsilon)\epsilon$, where
$D_0(\epsilon)$ is the density of states for noninteracting electrons;
$d^2=\frac{1}{4}(1-\frac{U}{U_c})$ with $U_c=8|K|$; and
$q=1-(\frac{U}{U_c})^2$. We thus recover a result of Ref.\
\cite{Kotliar:1986}.

In the strong-correlation limit, $U\gg t$, we find up to leading order
in $t/U$:  $d^2=4t^2/U^2$, $q=1-4t^2/U^2$, $K=-8t^2/U$, and
$m=1-8t^2/U^2$. Hence, the ground-state energy is $e=-4t^2/U$. This 
result is qualitatively correct: In the strong-correlation limit, the 
half-filled Hubbard model can be mapped onto the Heisenberg
antiferromagnet with an exchange coupling constant $J\equiv4t^2/U$. 
If we treat the electron spins as Ising spins and use that the
coordination number of our two-dimensional lattice is four, we also
obtain $e=-J$. Therfore, the mean-field theory gives rise to the
energy scale $J$ and correctly accounts for a ground-state energy of
the order of $-J$ per site. Finally, we note that the leading $1/U$
corrections do not depend on $t^{\prime}$. More generally, the entire
insulating phase is unaffected by next-nearest-neighbor hops. 
$t^{\prime}$ enters only via Eqs.\ (\ref{density}), 
(\ref{magnetization}) and (\ref{kinetic}). For the insulator, however,
only the band $\epsilon_{\vec{k}-}$ --- but the entire one ---
contributes to the $\vec{k}$ sums in these equations, and $t^{\prime}$
drops out.

The ground state for given model parameters $U$ and $\alpha\equiv
t^{\prime}/t$ corresponds to the minimum of the energy function
(\ref{Landau-function}). $\alpha$ is a measure for the degree of
magnetic frustration and is varied from zero to one. We may restrict
ourselves to positive $\alpha$, since a sign change of $\alpha$ is
tantamount to a particle-hole transformation,
$a_{i\eta}\rightarrow\exp\{i\vec{Q}\vec{R}_i\}a_{i|-\eta}^+$ 
($a$ and $a^+$ describe the quasiparticles with dispersion
(\ref{dispersions}), $\vec{Q}=(\pi,\pi)$, and $\vec{R}_i$ 
is the lattice vector to site $i$). The evolution of $e(m)$ as a
function of $U$ and $\alpha$ reveals how the transitions between the
various phases take place.

For $\alpha\ne0$, The inverse susceptibility, $\chi^{-1}$, is known
explicitly and changes its sign from positive to negative as $U$ is
increased to above $U_0\equiv8|K(0)|(\sqrt{|K(0)|/(|K(0)|
-t\gamma)}-1)$. Here, $1/\gamma\equiv(1/N)
\sum^{\prime}_{\vec{k}}1/|\cos k_x+\cos k_y|$ and
$\sum^{\prime}_{\vec{k}}$ denotes the sum over those regions of the
first Brillouin zone which are restricted by the condition
$\epsilon_{\vec{k}-}(X=0)<\tilde{\mu}(m=0)<\epsilon_{\vec{k}+}(X=0)$.
$U_0$ and $\gamma$ depend only on $\alpha$. 

Whether the system is metallic or insulating depends on the value of
the ground-state magnetization: If it exceeds a certain value,
$m_{\mbox{\scriptsize MIT}}$, a gap opens up in the single-particle
spectrum and the system goes insulating. This can be seen from Eq.\
(\ref{dispersions}) if we use that $m$ increases monotonically as a
function of $X$. Consequently, the insulator is always
antiferromagnetically ordered. We infer from Eqs.\
(\ref{dispersions})-(\ref{magnetization}) that $m_{\mbox{\scriptsize
MIT}}$ does not depend on $U$, but on the level of magnetic
frustration: Due to perfect nesting,
$\lim_{\alpha\rightarrow0}m_{\mbox{\scriptsize MIT}}=0$. As $\alpha$
is turned on, $m_{\mbox{\scriptsize MIT}}$ increases monotonically as
a function of $\alpha$.

\begin{center}
\begin{figure}
        \epsfig{file=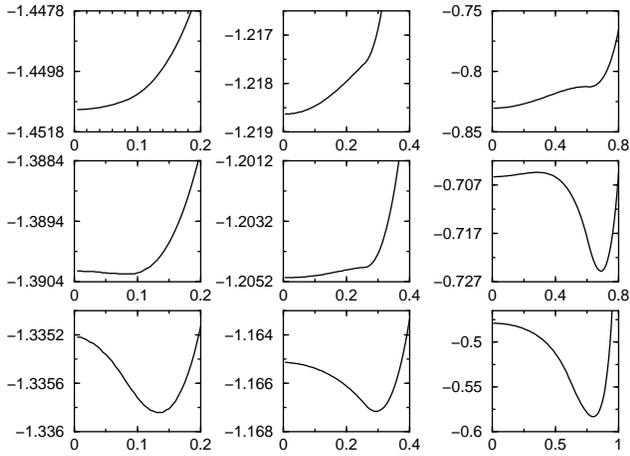,%
        width=3.3in}
        \vspace{6pt}
        \caption{The function $e(m)$ as a function of $U$ and the
magnetic frustration. The columns from left to right correspond to the
small-, intermediate-, and large-$\alpha$ regimes, respectively. The
respective values are $\alpha=0.02$, $0.1$, and $0.5$, and correspond
to $m_{\mbox{\scriptsize MIT}}=0.09$, $0.26$, and $0.61$,
respectively. Each column displays how $e(m)$ changes qualitatively
upon increasing $U$, from the paramagnetic metal (top row) to the
antiferromagnetic insulator (bottom row). The middle row shows
examples in the antiferromagnetic metal (first two plots), and one
after the metal-insulator transition has taken place but before the
local minimum at $m=0$ turns over into a local maximum (plot on the
right). The plot in the center has its minimum at a nonzero
magnetization, which is not discernible.}
\label{fig:energy_m}
\end{figure}
\end{center}

\begin{center}
\begin{figure}
        \epsfig{file=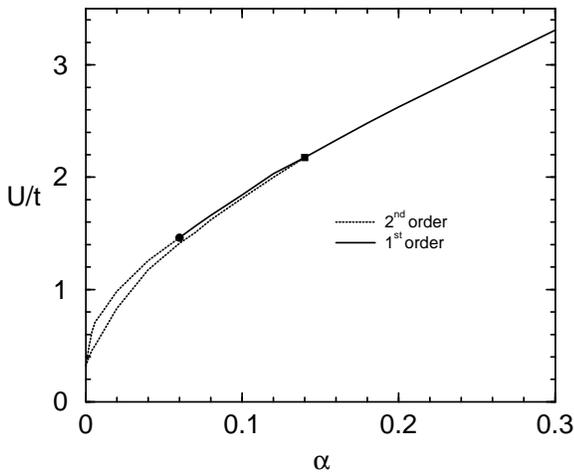,%
        width=3.3in}
        \vspace{6pt}
        \caption{The various phases and transitions as a function
of $\alpha\equiv t^{\prime}/t$ and $U$. The small-$U$ phase is the
paramagnetic metal, while the large-$U$ phase is the antiferromagnetic
insulator. In between, an antiferromagnetically ordered metallic phase
is sandwiched that disappears at a tricritical point (filled square). 
The dotted and filled lines indicate second- and first-order
transitions, respectively. The filled circle marks the point where
the metal-insulator transition changes its order.}
\label{fig:sb_phases}
\end{figure}
\end{center}

In our numerical investigation, we used a tetrahedron method
\cite{Gilat:1966} and, for $\alpha<0.2$ and $m<m_{\mbox{\scriptsize
MIT}}$, up to $10^4\times10^4$ points to do the $\vec{k}$ sums in
Eqs.\ (\ref{density}), (\ref{magnetization}) and (\ref{kinetic}). In
the discussion of our numerical results, we must distinguish between
three regimes of magnetic frustration and may restrict the discussion
of $e(m)$ to positive magnetizations. Fig.\ \ref{fig:energy_m}
illustrates how $e(m)$ evolves as a function of $U$ and $\alpha$. The
resulting phase diagram is displayed in Fig.\ \ref{fig:sb_phases}.

For {\em small} levels of magnetic frustration, $0<\alpha<0.06$
(first column of Fig.\ \ref{fig:energy_m}), we first find a
second-order transition from the paramagnetic to the antiferromagnetic
metal at a critical value $U_{\mbox{\scriptsize mag}}=U_0$. Upon
further increasing $U$, the resulting minimum is continuously shifted
towards higher magnetizations until it crosses $m_{\mbox{\scriptsize
MIT}}$ at a second critical value, $U_{\mbox{\scriptsize MIT}}$. 
Consequently, the metal-insulator transition is also of second
order. Since $m_{\mbox{\scriptsize MIT}}$ vanishes as
$\alpha\rightarrow0$, both transitions coincide in this limit,
$U_{\mbox{\scriptsize mag}}=U_{\mbox{\scriptsize MIT}}=U_0$. We also
know that $U_0=0$ if $\alpha=0$. 

For {\em intermediate} levels of magnetic frustration, 
$0.06\le\alpha<0.14$ (middle column of Fig.\ \ref{fig:energy_m}),
the magnetic transition is still of second order and occurs at $U_0$,
$U_{\mbox{\scriptsize mag}}=U_0$. On the other hand,
$m_{\mbox{\scriptsize MIT}}$ is sufficiently large for the
metal-insulator transition to take place differently: Before the
ground-state magnetization of the antiferromagnetic metal reaches
$m_{\mbox{\scriptsize MIT}}$ upon increasing $U$, a second 
minimum at a magnetization above $m_{\mbox{\scriptsize MIT}}$ has
emerged and become the absolute minimum of $e(m)$. Consequently, the
metal-insulator transition is now of first order. The transition lines
$U_{\mbox{\scriptsize MIT}}(\alpha)$ from the small-$\alpha$ and
intermediate-$\alpha$ regimes meet at $\alpha=0.06$ (filled circle in
Fig.\ \ref{fig:sb_phases}). At this point, the two degenerate minima
of the first-order transition merge at $m_{\mbox{\scriptsize MIT}}$. 
Finally, the antiferromagnetic metallic phase disappears gradually as
$\alpha\rightarrow0.14$ (filled square in Fig.\ \ref{fig:sb_phases}).

For {\em large} degrees of magnetic frustration, $\alpha\ge0.14$
(right column of Fig.\ \ref{fig:energy_m}), the antiferromagnetic and
metal-insulator transitions coincide, $U_{\mbox{\scriptsize
MIT}}=U_{\mbox{\scriptsize mag}}$, because the second-order transition
at $U_0$ is now preempted by the first-order one: By the time the
minimum at $m=0$ bifurcates, the one {\em above} $m_{\mbox{\scriptsize
MIT}}$ has already evolved into the absolute one, and remains to be
so, as $U$ is further increased.

We have also considered the Hubbard model in the Hartree-Fock
approximation. Earlier work on the $t-t^{\prime}$ Hubbard model in
Hartree-Fock approximation was carried out in Refs.\ \cite{Lin:1987},
\cite{Duffy:1997}, and \cite{Hofstetter:1998}. As for the
two-dimensional half-filled case, these works did not conclusively
answer whether an antiferromagnetic metal is stable in a certain
parameter regime.

\begin{center}
\begin{figure}
        \epsfig{file=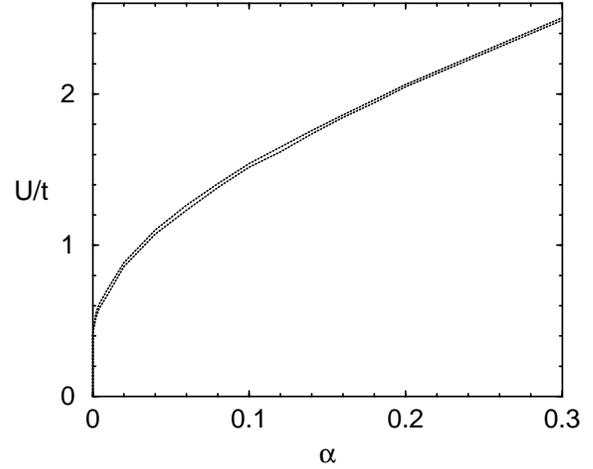,%
        width=3.3in}
        \vspace{6pt}
        \caption{The phase diagram in Hartree-Fock approximation: The
paramagnetic metal (small-$U$ regime) and antiferromagnetic insulator
(large-$U$ regime) are separated by a tiny range of $U$ values
corresponding to an antiferromagnetically ordered metal. Both
transitions are of second order.}
\label{fig:hf_phases}
\end{figure}
\end{center}

Within our formalism, the Hartree-Fock approximation turns out to be
tantamount to taking $q=1$ and $d^2(m)=(1-m^2)/4$ in Eq.\
(\ref{Landau-function}), while the function $K(m)$ is again determined
by Eqs.\ (\ref{density}), (\ref{magnetization}), and (\ref{kinetic}).
As a consequence, $\chi^{-1}=2(\gamma t-U/4)$, which changes its sign
at $U_0(\alpha)=4t\gamma(\alpha)$. This is the small-$\alpha$ limit of
the corresponding mean-field expression. Expanding
$e^{\mbox{\scriptsize HF}}(m)$ one step further yields the same
$-|m|^3\log|m|$ term as in mean-field theory, but with $q=1$. This
suggests that the magnetic transition is of second order and takes
place at $U_0(\alpha)$. In fact, our numerical investigation of the
evolution of $e^{\mbox{\scriptsize HF}}(m)$ as $U$ is increased
reveals the following behavior for all finite values of $\alpha$:
After a second-order transition from the paramagnetic to the
antiferromagnetic metal at $U_{\mbox{\scriptsize mag}}=U_0(\alpha)$,
the minimum of $e^{\mbox{\scriptsize HF}}(m)$ is rapidly but
continuously displaced towards higher magnetizations, until it exceeds
$m_{\mbox{\scriptsize MIT}}$. Thus, the metal-insulator-transition is
also of second order. Fig.\ \ref{fig:hf_phases} displays the phase
diagram in Hartree-Fock approximation. The error bars of the
transition lines $U_{\mbox{\scriptsize mag}}(\alpha)$ and
$U_{\mbox{\scriptsize MIT}}(\alpha)$ become of the same order than
$U_{\mbox{\scriptsize MIT}}(\alpha)-U_{\mbox{\scriptsize
mag}}(\alpha)$ for $\alpha\le0.001$, and our numerics is trustworthy
down to about $\alpha\sim0.002$. 

In summary, we have investigated the effect of magnetic frustration on
the metal-insulator transition in the two-dimensional half-filled
Hubbard model within a slave-boson approach, and we have compared our
results to the Hartree-Fock approximation. Within the slave-boson
mean-field theory, our main results are: First, magnetic frustration
helps stabilize an antiferromagnetic metal for $t^{\prime}/t\le0.14$.
Second, for $t^{\prime}/t\ge0.06$, the metal-insulator transition is
of first order. Finally, all other transitions between the various
phases are of second order. By contrast, in Hartree-Fock
approximation, the magnetic and metal-insulator transitions are always
separate and of second order. Both in Hartree-Fock approximation and
in the slave-boson mean-field theory, the insulator is always
antiferromagnetically ordered.

We gratefully acknowledge discussions with R. Fr\'esard and
R. Chitra. This work was supported by the NSF DMR 95-29138. E.L. is
partly funded by the Deutsche Forschungsgemeinschaft.  


\newpage


\vspace{-6pt}

\begin{thebibliography}{10}
\vspace{-6pt}
\bibitem{Mott:1961}
N.~F. Mott, Philos. Mag. {\bf 6},  287  (1961).

\bibitem{Carter:1993}
S.~A. Carter {\it et~al.}, Phys. Rev. B {\bf 48},  16841  (1993).

\bibitem{Imada:1998}
M. Imada, A. Fujimori, and Y. Tokura, Rev. Mod. Phys. {\bf 70},  1039
(1998).

\bibitem{Gautier:1975}
F. Gautier {\it et~al.}, Phys. Lett. A {\bf 53},  31  (1975).

\bibitem{Sudo:1992}
S. Sudo, J. Magn. Magn. Mater. {\bf 114},  57  (1992).

\bibitem{Brinkman:1970}
W.~F. Brinkman and T.~M. Rice, Phys. Rev. B {\bf 2},  4302  (1970).

\bibitem{Gutzwiller:1963}
M.~C. Gutzwiller, Phys. Rev. Lett. {\bf 10},  159  (1963).

\bibitem{Gutzwiller:1965}
M.~C. Gutzwiller, Phys. Rev. {\bf 137},  A1726  (1965).

\bibitem{Rozenberg:1994}
M. Rozenberg, G. Kotliar, and X. Zhang, Phys. Rev. B {\bf 49},  10181
(1994).

\bibitem{Lin:1987}
H. Lin and J. Hirsch, Phys. Rev. B {\bf 35},  3359  (1987).

\bibitem{Duffy:1997}
D. Duffy and A. Moreo, Phys. Rev. B {\bf 55},  676  (1997).

\bibitem{Chitra:1998}
R. Chitra and G. Kotliar, cond-mat/9811144.

\bibitem{Kotliar:1986}
G. Kotliar and A.~E. Ruckenstein, Phys. Rev. Lett. {\bf 57},  1362
(1986).

\bibitem{Lilly:1990}
L. Lilly, A. Muramatsu, and W. Hanke, Phys. Rev. Lett. {\bf 65},  1379
(1990).

\bibitem{Fresard:1991}
R. Fr\'esard, M. Dzierzawa, and P. W\"olfle,
Europhys. Lett. {\bf 15},  325  (1991).

\bibitem{Fresard:1992}
R. Fr\'esard and P. W\"olfle, J. Phys. Condensed Matter {\bf 4},
3625  (1992).

\bibitem{Moeller:1993}
B. M\"oller, K. Doll, and R. Fr\'esard, J. Phys. Condensed Matter {\bf
5},  4847  (1993).

\bibitem{Seibold:1998}
S. Seibold and V. Hizhnyakov, Phys. Rev. B {\bf 57},  6937  (1998).

\bibitem{Gilat:1966}
G. Gilat and L.~J. Raubenheimer, Phys. Rev. {\bf 144},  390  (1966).

\bibitem{Hofstetter:1998}
W. Hofstetter and D. Vollhardt, Ann. Physik {\bf 7}, 48 (1998).

\end{thebibliography}
\end{document}